\documentclass[article,longbibliography,nofootinbib,superscriptaddress]{revtex4-2}
\usepackage{graphicx}
\usepackage{xcolor}
\usepackage{tikz}
\usepackage{amsmath}
\usepackage{booktabs}
\usepackage{enumitem}
\usepackage[colorlinks=true, urlcolor=blue, linkcolor=black, citecolor=black]{hyperref}
\usetikzlibrary{arrows.meta, positioning, shapes.geometric,
                shapes.arrows, calc, backgrounds, fit,
                decorations.pathreplacing}

\usepackage[normalem]{ulem}
\usepackage{fancyhdr}

\definecolor{prlblue}{RGB}{31,78,121}
\definecolor{prlred}{RGB}{153,31,35}
\definecolor{prlgray}{RGB}{68,68,68}
\definecolor{prllg}{RGB}{120,120,120}
\definecolor{prlbluetint}{RGB}{234,240,247}
\definecolor{prlredtint}{RGB}{248,234,233}
\fancypagestyle{plain}{%
  \fancyhf{}%
}

\begin{document}

\title{From Particles to Policy: \\ Technical Building Blocks for Multi-State SAI Coordination}

\author{Roby Yahav}
\thanks{\href{mailto:r.yahav@stardust-initiative.com}{r.yahav@stardust-initiative.com}}
\affiliation{Stardust Labs, Ness Ziona, Israel; \url{https://www.stardust-initiative.com}}
\author{Amyad Spector}
\thanks{\href{mailto:a.spector@stardust-initiative.com}{a.spector@stardust-initiative.com}}
\affiliation{Stardust Labs, Ness Ziona, Israel; \url{https://www.stardust-initiative.com}}
\author{Doron Kushnir}
\thanks{\href{mailto:doron.k@stardust-initiative.com}{doron.k@stardust-initiative.com}}
\affiliation{Stardust Labs, Ness Ziona, Israel; \url{https://www.stardust-initiative.com}}
\author{Matthew C. Waxman}
\thanks{Liviu Librescu Professor of Law, \href{mailto:matthew.waxman@law.columbia.edu}{matthew.waxman@law.columbia.edu}}
\affiliation{Columbia University, New York, USA}
\date{May 2026}

\begin{abstract}
Stratospheric aerosol injection (SAI) is a solar radiation modification technique, proposed as an interim measure to offset warming while greenhouse gas (GHG) emissions are reduced. This paper discusses a possible SAI implementation route - an alternative to sulfate aerosols formed in situ - based on engineered solid particles having dedicated properties such as size, composition, surface chemistry, and traceable origin, supporting safety, controllability, and functionality needed for SAI systems. These engineered properties also open up options for any future multi-state coordination of SAI through two technical building blocks: (1) the SAI-induced radiative forcing (SRF) - the magnitude of the cooling effect attributable specifically to the SAI layer - as an operator-independent quantity, derivable from direct aerosol-layer measurements; and (2) particle traceability through identifying signatures embedded at production. Both could feed into a shared, publicly accessible monitoring database open to independent interrogation, addressing several governance challenges by anchoring compliance assessments in measurable parameters. Drawing on precedents from the Montreal Protocol, IAEA safeguards, and other regimes, we show that shared technical metrics have historically enabled multi-state cooperation, and we argue the same could apply to SAI. We describe a phased pathway in which the technical capabilities and coordination practices that would use them are developed and tested together, at scales orders of magnitude below operational deployment. To be clear - we regard SAI deployment as premature; the conditions under which it might be considered have not been met. The paper does not propose a governance framework; rather, it identifies technical infrastructure that could support a wide range of such frameworks.

\vspace{1.5em}

\noindent\textbf{Plain-language summary.}
Stratospheric aerosol injection (SAI) - adding reflective particles into the upper atmosphere - has been proposed as a way to temporarily cool the planet while greenhouse gas emissions are reduced. This paper does not argue that SAI should be deployed or that it is ready for that. It identifies two features of the engineered particle layer - how much sunlight the layer reflects back to space, and the exact identity of each particle - that would give cooperating states a shared, observable basis for coordinating any such climate stabilization program. Particle identity enables tracing every contribution to the layer, whether from participating states, outside actors, or gaps in compliance. Both features depend on using purpose-built solid particles, not sulfates. Those same measurements could ground questions of responsibility, verification, and dispute resolution in observed facts rather than disputed estimates of climate impacts. This kind of measurement-first approach has supported international cooperation on hard problems before, from nuclear arms control to ozone protection. The paper also outlines a step-by-step pathway in which the technical tools and the governance practices around them are built and tested together, well before SAI itself could possibly be deployed.
\end{abstract}

\maketitle
\thispagestyle{fancy}

\section{Introduction}
\label{sec:introduction}

Stratospheric aerosol injection (SAI) is a solar radiation modification (SRM) technique, proposed as an interim measure to offset warming while greenhouse gas (GHG) emissions are reduced \cite{RoyalSociety2025,UNEP2023}. It involves introducing reflective aerosols into the lower stratosphere, roughly around 20 km altitude, where they would scatter a small fraction of incoming sunlight back to space, thereby cooling the lower atmosphere and Earth's surface below. One approach is to form sulfate aerosols in situ from injected precursor gases, partly mimicking the transient global cooling observed after large volcanic eruptions such as Mount Pinatubo in 1991. An alternative approach, on which this paper focuses, is to disperse engineered solid particles whose size, composition, surface chemistry, and traceable origin can be specified at production - capabilities explored in an accompanying body of technical work, e.g., \cite{Waxman2026,Stardust2026Composite}. In either case, sustained cooling would require repeated injection for decades. Key physical parameters related to SAI particle measurements are summarized in Annex~A. The present paper builds on the engineered-particle work cited above by examining how these particle properties could enable future multi-state coordination of SAI and address some often-cited governance challenges.

Beyond the technical barriers, any sustained SAI program would face political ones. Should one ever be pursued at scale, it would most plausibly operate within a multi-state coordination framework in which states jointly authorize, finance, and oversee the program. A unilateral path - a single sufficiently powerful state acting alone - is technically conceivable \cite{Horton2025_WhoDeploy} but unlikely to be politically durable across the multi-decadal time horizon SAI would require: as Horton observes, ``SAI geoengineering is ruled by a `logic of multilateralism''' \cite{Horton2011}. There are many possible multilateral frameworks, and in general, states could coordinate execution either across several national operators, each accountable to its own government, or by delegating to a single, jointly mandated entity. They would face problems comparable to those the international community has confronted with meaningful (if partial) success - in managing ozone depletion, arms control, and the shared use of outer space.

To be clear, we regard the possibility of SAI deployment as premature; the conditions under which governments might consider large-scale SAI deployment have not yet been met. Moreover, this paper does not propose a governance framework, nor does it attempt to identify the comprehensive set of issues any such framework would have to resolve \footnote{Like Biniaz and Bodansky \cite{Biniaz2025}, and consistent with the polycentric pattern by which complex governance arrangements have historically emerged \cite{Ostrom2010}, we expect that any coordinated framework - if and when it ever becomes both necessary and practical - would most likely emerge incrementally, through norm-setting built outward from smaller, more localized arrangements, rather than negotiated top-down. That said, the purpose of this paper is not to propose any specific approach.}. Rather, building on the body of technical work cited above, we describe two capabilities that engineered solid particles enable - both relating to what can be measured, by whom, and with what level of independence from the operator - and we sketch how those capabilities could open up options and support a wide range of possible coordination or governance approaches. The two capabilities are: (1) the time- and space-dependent SAI-induced radiative forcing (SRF) - the magnitude of the cooling effect attributable specifically to the SAI layer - as an operator-independent quantity, derivable from direct aerosol-layer measurements; and (2) particle traceability - identifiable signatures embedded in the particles themselves - which makes the origin of any sampled particle determinable even after stratospheric mixing. Both could feed into a publicly accessible monitoring database open to independent interrogation. Together, they would commit to a declared SAI program that is observable rather than reliant on operator self-report, and they would make the source of any deviation determinable from physical evidence.

A substantial and growing literature addresses what principles future SAI governance should embody, including accountability, transparency, equity, and inclusive engagement \cite{NAS2021,UNEP2023,RoyalSociety2025,COC2023,Reynolds2019,Jinnah2019,NAS2020}. A related set of practical challenges is also widely recognized - for instance, reconciling national interests over cooling targets \cite{MorenoCruz2012,Ricke2013}, distinguishing the contributions of different actors after stratospheric mixing \cite{Horton2015,Gerrard2018,Martin2025}, and maintaining operational continuity across decades \cite{Parker2018,Jones2013,McCusker2014}. The capabilities described here are intended as a technical input to that conversation - specifically, on how compliance with declared SAI commitments could be independently observed in practice. Without a measurement infrastructure that makes compliance auditable, commitments reduce to trust - and in a domain as consequential as SAI, trust alone is insufficient.

Crucially, both capabilities can be prototyped at scales orders of magnitude below operational deployment, so willing states and research groups could begin testing them - and the institutional practices that would use them - long before any decision about deployment is on the table.

Section~\ref{sec:building-blocks} describes the two technical capabilities in detail. Section~\ref{sec:challenges} shows how these building blocks could support multi-state coordination, sketching several illustrative challenges they would help to address. Section~\ref{sec:precedents} draws lessons from existing multilateral regimes that have used measurement-anchored cooperation to manage contested issues. Section~\ref{sec:phased} outlines a phased pathway for developing the technical capabilities and rehearsing the multi-state coordination practices that could use them. Section~\ref{sec:scenarios} illustrates some practical scenarios where these technical capabilities could be useful. Section~\ref{sec:conclusion} concludes. Annex~A summarizes the underlying physical parameters.

\section{Two Technical Building Blocks}
\label{sec:building-blocks}

This section describes the two technical capabilities introduced in Section~\ref{sec:introduction}: SAI-induced radiative forcing (SRF) as an operator-independent quantity, derivable from direct aerosol-layer measurements, and particle traceability through identifying signatures embedded at production. We describe these capabilities first, before turning in Section~\ref{sec:challenges} to the practical multi-state coordination problems where they could be useful.

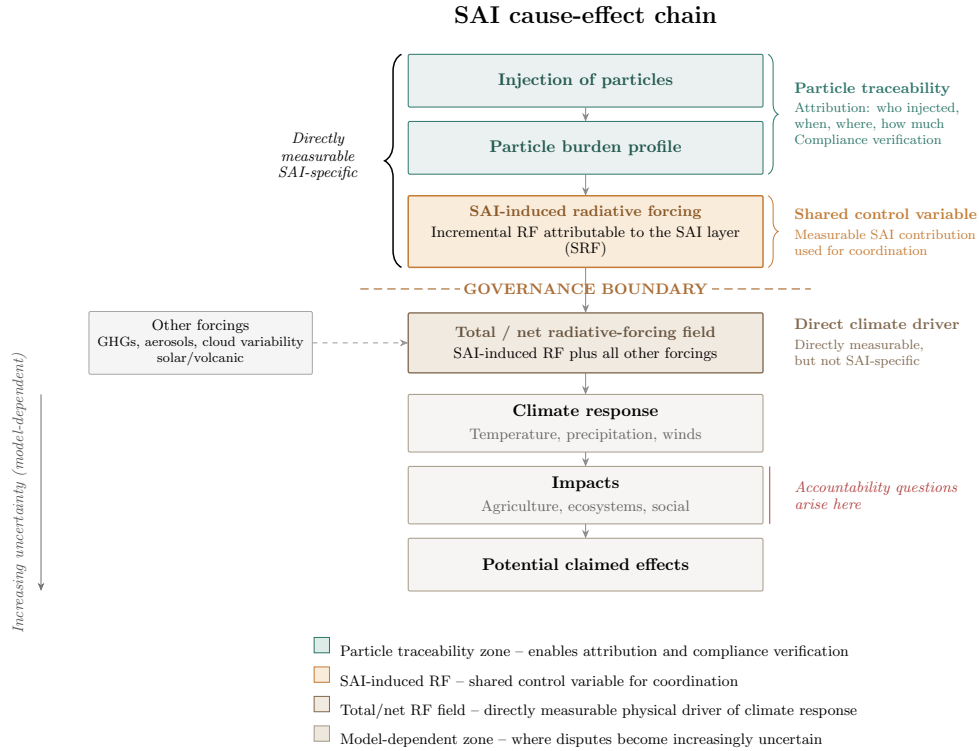
\begin{figure*}[ht]
\centering

\definecolor{tracecolor}{RGB}{47,119,107}
\definecolor{tracetint} {RGB}{222,236,231}
\definecolor{rfcolor}   {RGB}{196,125,55}
\definecolor{rftint}    {RGB}{250,236,219}
\definecolor{netcolor}  {RGB}{130,105,80}
\definecolor{nettint}   {RGB}{241,234,224}
\definecolor{modelcolor}{RGB}{166,155,138}
\definecolor{modeltint} {RGB}{237,231,222}
\definecolor{liabred}   {RGB}{180,70,70}
\definecolor{governance}{RGB}{170,110,55}

\newcommand{\legendswatch}[2]{%
  \tikz[baseline=-0.5ex]{\fill[#2] (0,0) rectangle (0.32,0.32);
                          \draw[#1, line width=0.6pt] (0,0) rectangle (0.32,0.32);}%
}

\resizebox{0.72\textwidth}{!}{%
\begin{tikzpicture}[
    >={Stealth[length=5pt, width=4pt]},
    node distance=0.30cm,
    every node/.style={outer sep=0pt},
    mainbox/.style={
        rectangle, rounded corners=1pt,
        draw=#1, line width=0.6pt,
        fill=#1!10,
        minimum width=7.4cm, minimum height=1.1cm,
        align=center, inner sep=6pt,
        font=\normalsize,
    },
    rfbox/.style={
        rectangle, rounded corners=1pt,
        draw=rfcolor, line width=0.9pt,
        fill=rftint,
        minimum width=7.4cm, minimum height=1.35cm,
        align=center, inner sep=6pt,
    },
    netbox/.style={
        rectangle, rounded corners=1pt,
        draw=netcolor, line width=0.8pt,
        fill=nettint,
        minimum width=7.4cm, minimum height=1.25cm,
        align=center, inner sep=6pt,
    },
    sourcebox/.style={
        rectangle, rounded corners=1pt,
        draw=black!35, line width=0.5pt,
        fill=black!4,
        minimum width=3.0cm, minimum height=0.9cm,
        align=center, inner sep=5pt,
        font=\small,
    },
    brlbl/.style={
        font=\small,
        align=left,
    },
    flowarrow/.style={
        -{Stealth[length=5pt, width=4pt]},
        line width=0.5pt, black!45,
    },
    sidearrow/.style={
        -{Stealth[length=5pt, width=4pt]},
        line width=0.5pt, black!40,
        dashed,
    },
]

\node[font=\Large\bfseries] (title) at (0,0) {SAI cause-effect chain};

\node[mainbox=tracecolor, below=0.55cm of title] (inject)
    {\textbf{\textcolor{tracecolor!75!black}{Injection of particles}}};

\node[mainbox=tracecolor, below=0.30cm of inject] (burden)
    {\textbf{\textcolor{tracecolor!75!black}{Particle burden profile}}};

\draw[flowarrow] (inject.south) -- (burden.north);

\draw[decorate, decoration={brace, amplitude=6pt, raise=3pt},
      line width=0.6pt, tracecolor]
    (inject.north east) -- (burden.south east);

\node[brlbl, text=tracecolor,
      right=0.55cm of $(inject.east)!0.5!(burden.east)$] {%
    \textbf{Particle traceability}\\[1pt]
    {\footnotesize Attribution: who injected,}\\[-1pt]
    {\footnotesize when, where, how much}\\[-1pt]
    {\footnotesize Compliance verification}%
};

\node[rfbox, below=0.45cm of burden] (sairf) {%
    \textbf{\textcolor{rfcolor!70!black}{SAI-induced radiative forcing}}\\[2pt]
    {\small Incremental RF attributable to the SAI layer}\\[-1pt]
    {\small (SRF)}%
};

\draw[flowarrow] (burden.south) -- (sairf.north);

\draw[decorate, decoration={brace, amplitude=6pt, raise=3pt},
      line width=0.6pt, rfcolor]
    (sairf.north east) -- (sairf.south east);

\node[brlbl, text=rfcolor, right=0.55cm of sairf.east] {%
    \textbf{Shared control variable}\\[1pt]
    {\footnotesize Measurable SAI contribution}\\[-1pt]
    {\footnotesize used for coordination}%
};

\draw[decorate,
      decoration={brace, amplitude=10pt, raise=4pt},
      line width=0.9pt, black]
    (sairf.south west) -- (inject.north west);

\node[font=\small\itshape, text=black, align=center,
      left=0.95cm of $(inject.west)!0.5!(sairf.west)$] {%
    Directly\\measurable\\[-1pt]
    SAI-specific};

\coordinate (gbL) at ([xshift=-1.0cm, yshift=-0.45cm]sairf.south west);
\coordinate (gbR) at ([xshift= 1.0cm, yshift=-0.45cm]sairf.south east);

\draw[dashed, line width=0.7pt, governance,
      dash pattern=on 5pt off 3pt] (gbL) -- (gbR);

\node[font=\small\bfseries, text=governance, fill=white, inner sep=2pt]
      at ($(gbL)!0.5!(gbR)$) {GOVERNANCE BOUNDARY};

\node[netbox, below=0.95cm of sairf] (netrf) {%
    \textbf{\textcolor{netcolor!80!black}{Total / net radiative-forcing field}}\\[2pt]
    {\small SAI-induced RF plus all other forcings}%
};

\draw[flowarrow] (sairf.south) -- (netrf.north);

\node[sourcebox, left=2.0cm of netrf] (otherf) {%
    Other forcings\\[-1pt]
    {\footnotesize GHGs, aerosols, cloud variability}\\[-1pt]
    {\footnotesize solar/volcanic}%
};

\draw[sidearrow] (otherf.east) -- (netrf.west);

\node[brlbl, text=netcolor, right=0.55cm of netrf.east] {%
    \textbf{Direct climate driver}\\[1pt]
    {\footnotesize Directly measurable,}\\[-1pt]
    {\footnotesize but not SAI-specific}%
};

\node[mainbox=modelcolor, below=0.45cm of netrf] (climate) {%
    \textbf{Climate response}\\[2pt]
    {\small\color{black!60} Temperature, precipitation, winds}%
};

\node[mainbox=modelcolor, below=0.30cm of climate] (impacts) {%
    \textbf{Impacts}\\[2pt]
    {\small\color{black!60} Agriculture, ecosystems, social}%
};

\node[mainbox=modelcolor, below=0.30cm of impacts] (claims) {%
    \textbf{Potential claimed effects}%
};

\draw[flowarrow] (netrf.south)   -- (climate.north);
\draw[flowarrow] (climate.south) -- (impacts.north);
\draw[flowarrow] (impacts.south) -- (claims.north);

\draw[line width=0.6pt, liabred]
    ([xshift=4pt]impacts.north east) -- ([xshift=4pt]impacts.south east);

\node[brlbl, text=liabred, right=0.55cm of impacts.east,
      font=\small\itshape] {Accountability questions\\arise here};

\coordinate (arrTop) at ([xshift=-1.0cm]otherf.west |- climate.north);
\coordinate (arrBot) at ([xshift=-1.0cm]otherf.west |- claims.south);

\draw[-{Stealth[length=5pt, width=4pt]}, line width=0.6pt, black!55]
    (arrTop) -- (arrBot);

\node[font=\small\itshape, text=black!65, align=center, rotate=90]
      at ([xshift=-0.45cm]$(arrTop)!0.5!(arrBot)$)
      {Increasing uncertainty (model-dependent)};

\node[below=0.85cm of claims.south, anchor=north,
      font=\small, align=left] (legend) {%
\begin{tabular}{@{}l@{\hspace{6pt}}l@{}}
\legendswatch{tracecolor}{tracetint} &
  Particle traceability zone -- enables attribution and compliance verification\\[3pt]
\legendswatch{rfcolor}{rftint} &
  SAI-induced RF -- shared control variable for coordination\\[3pt]
\legendswatch{netcolor}{nettint} &
  Total/net RF field -- directly measurable physical driver of climate response\\[3pt]
\legendswatch{modelcolor}{modeltint} &
  Model-dependent zone -- where disputes become increasingly uncertain\\
\end{tabular}%
};

\node[below=0.35cm of legend.south, font=\footnotesize\itshape,
      text=black!55]
  {Adapted from Fuglestvedt et al.\ (2003) and Waxman et al.\ (2026).};

\end{tikzpicture}}

\caption{Cause-effect chain from SAI injection to downstream effects, adapted from Fuglestvedt et al.~\cite{Fuglestvedt2003} and Waxman et al.~\cite{Waxman2026}. Above the governance boundary, the chain contains the SAI-specific quantities most relevant to multi-state coordination: particle injection, particle burden, and the SAI-induced radiative forcing. The SAI-induced radiative forcing provides the shared control variable because it isolates the incremental contribution attributable to the SAI layer. Below the boundary, this SAI-specific forcing combines with GHG, background aerosols, solar and volcanic forcing, and other forcings to form the total or net radiative-forcing field. That total forcing field is directly measurable and is the physical driver of climate response, but it is not itself SAI-specific. The subsequent mapping from forcing to climate response, impacts, and claimed effects is increasingly model-dependent. See Annex~A on the engineering basis for particle tagging, enabling attribution of stratospheric particles to production batches.}
\label{fig:cause_effect_chain}
\end{figure*}

\noindent\textbf{SAI-induced radiative forcing as the shared control variable.}
The international community is already, in effect, trying to govern radiative forcing (RF)\footnote{Radiative forcing is the change in the net energy flux at the top of the atmosphere (TOA) caused by an external driver such as GHG or aerosols, measured in power per square area. Positive forcing warms the surface; negative forcing, such as the incremental forcing produced by an SAI layer, cools it. In this paper, we distinguish between the total or net RF field and the SAI-induced radiative forcing. The latter, denoted hereafter as SRF, is the incremental RF attributable specifically to the SAI layer. The total or net RF field also includes GHG, background aerosols, solar and volcanic forcing, and other natural and anthropogenic drivers.}: every GHG mitigation target, every emissions budget, and every net-zero pledge is an indirect attempt to control how much radiation the atmosphere traps. SAI makes one component of this control problem explicit and direct: the incremental radiative forcing attributable to the SAI layer, i.e., the direct effect of the SAI aerosol burden. We refer to this quantity as SAI-induced radiative forcing, or SRF.

Figure~\ref{fig:cause_effect_chain} illustrates the cause-effect chain from injection downstream. For multi-state SAI coordination, the natural primary control variable is the allowed envelope of SRF, rather than the full total RF field. Auditing compliance would then turn on whether the SAI-attributable forcing remains within the agreed SRF envelope, rather than on tracing any specific weather or climate event to SAI.

Operationally, SRF would be monitored primarily by measuring the aerosol layer itself, rather than attempting to detect SRF directly as a small residual in the full TOA radiation budget. The controlled object is the SAI aerosol burden: its mass, altitude, latitude distribution, and size distribution. From these observables, SRF can be inferred with radiative-transfer calculations whose inputs are directly measured and whose uncertainties can be calibrated through laboratory and field measurements. Direct TOA measurements remain essential as an independent large-scale consistency check, but they are less suitable as a primary compliance variable: they measure the combined radiation field from all forcings, and they require extracting a small SAI perturbation from a large, variable background. Measuring the aerosol layer is also substantially simpler and less expensive than measuring a small residual in the full TOA radiation budget. Annex~A provides the quantitative basis for this hierarchy.

Below this governance boundary, SRF combines with GHG, background aerosols, solar and volcanic forcing, cloud-related variability, and other drivers to form the total or net radiative environment that governs the climate response. Moving further down the chain, from total forcing to climate response, impacts, and claimed effects, policy relevance increases but so does model dependence. In this hierarchy, SRF is the highest-level quantity that is at once measurable, SAI-specific, and policy-relevant; the subsequent mapping from forcing to downstream effects is where attribution and modeling uncertainty increase.

Concretely, we propose using time- and space-dependent SRF as the shared control variable. Building this monitoring capability - space-based instruments, stratospheric sampling platforms, and qualified analytical laboratories - is itself a substantial multilateral undertaking; observables and methods are detailed in Annex~A. The resulting data, covering particles, SRF, and climate observables, must be pooled in a shared, publicly accessible monitoring database open to independent interrogation. Such a database would be central to making compliance auditable, and would also satisfy international law's notification and consultation duties for transboundary activities \cite{Reynolds2021}.

Because SRF can be independently estimated by any state with appropriate instruments, measurements can then be compared across states against a common physical quantity and a common uncertainty bound. This reduction in measurement uncertainty is what makes SRF a viable \emph{coordination} variable, not merely a scientific one: the tighter the uncertainty bounds, the more credibly the coordination framework can distinguish compliant behavior from deviation.

\noindent\textbf{Particle traceability for attribution.}
The second capability is encoding identifying information in the particles themselves. If particles carry verifiable origin signatures, the coordination framework gains the ability to verify compliance, detect unauthorized injection, and attribute forcing deviations to their source, even after stratospheric mixing\footnote{The term \emph{mixing} refers to stratospheric transport processes that disperse and blend aerosols from distinct sources, so that routine measurements of the aerosol field cannot be decomposed into individual actors' SAI contributions.} has blended aerosols from multiple contributors. The concept has two layers. First, a declared batch identifier - based on environmentally benign stable trace elements or isotope ratios deliberately embedded during manufacture - allows any party with appropriate sampling equipment to match observed particles to declared production and injection records. Second, a manufacturing-origin signature, arising from raw materials, process conditions, equipment, or deliberately inserted factory-specific trace markers, can be known only to the coordinating authority and accredited laboratories. This hidden layer verifies that particles carrying a declared tag were actually made in an authorized facility; an actor copying the public batch identifier would still be detected if the factory-origin signature did not match the reference library. Because these signatures are intrinsic to the particle, stripping or spoofing them is, by design, intended to require altering the aerosol's compositional properties in detectable ways (see Annex~A for tagging-scheme considerations).

A relevant institutional analogy is the International Atomic Energy Agency (IAEA) safeguards regime, which has established a now-familiar pattern: production-to-use tracking via unique identifiers, verified chains of custody, and independent audit, with discrepancies triggering investigation. It is this pattern of continuous multilaterally-agreed accountancy and inspection that we propose to adapt to SAI, where the particle itself becomes the unit of account.

The two building blocks could apply to both centralized and distributed operational architectures. In a multi-operator architecture, SAI-induced radiative forcing measurements could verify whether the coalition collectively stays within agreed parameters; particle traceability could then answer the question that SRF alone cannot: which operators' outputs make up the aggregate signal. This would matter first in the routine case of overt multi-operator activity, where attribution is needed for cost-sharing, performance accounting, and identifying the source of any over- or under-delivery against plan. The same capability would also cover the harder case of covert activity by a non-participant or by a participant exceeding its agreed allocation: the same chemical and isotopic signatures that enable routine attribution also allow detection of material whose signature does not match any declared program. In a single-operator architecture, SRF monitoring would become the primary means by which member states verify that the authorized operator is faithfully executing the joint mandate, and traceability would shift from inter-member attribution toward operator auditability - verifying that observed particles belong to the authorized program, distinguishing authorized from unauthorized injection, and auditing execution at the level of batch, platform, place, and time. In both architectures, the external-attribution function - identifying particles that lack any authorized signature - would remain essential for detecting rogue actors' activity (this may require additional detection and forensic capabilities to be developed).

\begin{figure}[!htbp]
\centering
\includegraphics[width=\linewidth]{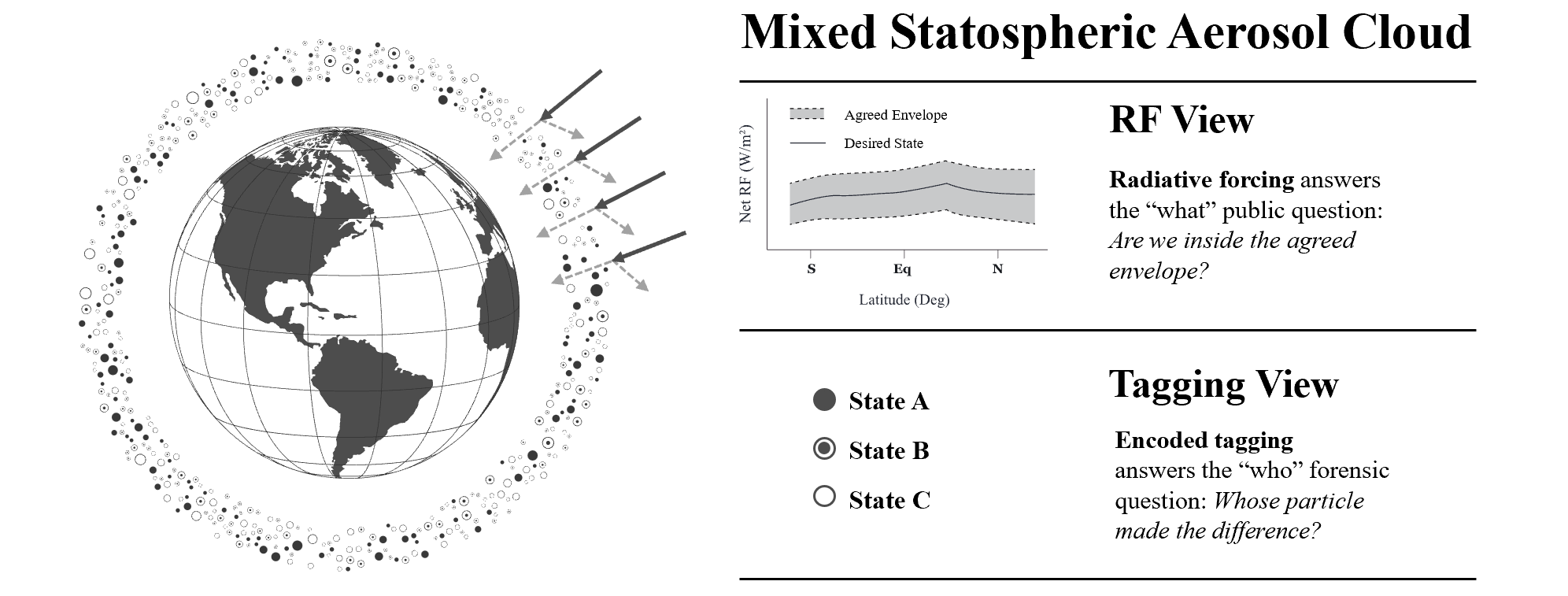}
\caption{Both building blocks applied to the same physical reality - a mixed stratospheric aerosol cloud containing tagged particles from multiple states. The SRF view measures aggregate SAI-induced radiative forcing to answer the compliance question: are we inside the agreed envelope? The tagging view reads particle signatures to answer the forensic question: whose particle made the difference?}
\label{fig:mixed_cloud}
\end{figure}

\section{How the Building Blocks Could Support Multi-State SAI Coordination}
\label{sec:challenges}

With the technical capabilities established, we sketch several governance challenges that any multi-state SAI arrangement would face. This is not a full catalog of such challenges, nor a claim that we resolve any of them fully. Rather, we focus on cases where the building blocks introduced in Section~\ref{sec:building-blocks} could provide the most direct support \cite{Bodansky2024}.

\noindent\textbf{Reconciling national interests.}
States would start from different positions on cooling targets, latitudinal distribution, cost-sharing, governance structure, termination conditions, and the relationship between SAI and emissions reduction \cite{Reynolds2019,Weitzman2015,Ricke2013}. A tropical state facing imminent monsoon disruption would weigh different forcing profiles than an arctic nation concerned about permafrost; a major emitter will weigh mitigation commitments differently than a vulnerable island state. The goal of any coordination framework would not be to eliminate these starting differences but to develop approaches whose benefits converge enough across participating states to make agreement possible. Technical infrastructure alone cannot do this work, but SRF as a shared control variable would convert the question into a negotiable forcing-envelope schedule. Because SRF can be independently measured from space and at the surface - subject to the precision and averaging-scale assumptions discussed in Annex~A - every state could verify commitments using its own instruments, and particle traceability enables attribution when deviations occur.

\noindent\textbf{Ongoing verification.}
Any participant could quietly adjust injection rates, and stratospheric mixing would obscure origins within weeks, leaving no convenient accounting trail \cite{Horton2015,Gerrard2018,Butchart2014}. Continuous SRF monitoring could detect deviations in near-real time, while particle traceability could identify the responsible actor even after mixing. Both functions would depend on the shared monitoring database introduced in Section~\ref{sec:building-blocks} - without independent access to data, compliance verification would reduce to relying on self-reports.

\noindent\textbf{Distinguishing contributions.}
After weeks of stratospheric transport, particles from multiple sources would blend together \cite{Gerrard2018}, making it difficult to determine which actor contributed what to the resulting aerosol layer. Particle traceability would address this directly. Sampled particles from coalition members would carry their declared origin signatures regardless of mixing, allowing each member's contribution to be identified after the fact. A non-member's untagged or differently-tagged particles would be equally identifiable through routine stratospheric sampling, without requiring any cooperation from the non-member \cite{Martin2025}. Combined with SRF monitoring - which quantifies the magnitude of any deviation - the coordination framework could identify both the responsible party and the scale of any discrepancy from declared production, whether the source is internal to the coalition or external.

\noindent\textbf{Attribution and auditing.}
The building blocks together could support a basic audit function: SRF monitoring would quantify deviations from declared envelopes, and particle traceability would identify the responsible actor. The question would become observable - did the operator maintain SRF within its committed envelope? - rather than contested - did some specific event flow from a state's actions? An observed deviation would establish a factual finding grounded in physical evidence. How participating states use such findings is outside the scope of this paper.

\noindent\textbf{Operational continuity and termination risk.}
Even with full compliance, operational failures over decades would be inevitable: aircraft groundings, supply-chain disruptions, volcanic eruptions, or political crises could all interrupt injection schedules. If SAI were abruptly halted, the masking effect would disappear within one to three years, while accumulated GHG forcing would persist, producing a termination shock that could overwhelm the adaptive capacity of ecosystems and societies \cite{Parker2018,Jones2013,McCusker2014}. SRF monitoring would provide early warning when injection rates drop, giving a response window of order a year before surface temperatures respond. Particle tagging would identify which actor has ceased injection and by how much, enabling the coordination framework to distinguish mechanical failure from deliberate withdrawal and to activate pre-positioned reserves accordingly.

\noindent\textbf{A note on single-operator architectures.}
Under a single-operator architecture - several states might jointly authorize and oversee, but execution might be delegated to a single mandated operator - the same challenges would arise in altered form (e.g., compliance would become a principal-agent question, and the technical detection of non-member contributions would shift to detecting under-funding by member states). The technical building blocks could apply equally to both architectures; how each challenge would manifest in single-operator execution is illustrated in Section~\ref{sec:scenarios}.

These coordination challenges are not unique to SAI: other contested international regimes - ozone depletion, nuclear proliferation, spectrum use, arms control, maritime and aviation tracking - faced analogous problems of reconciling interests, verification, distinguishing contributions, and continuity. In each case, progress became possible not only through diplomatic ingenuity but because a shared, independently measurable technical variable made compliance observable across adversarial states. The next section examines how such metrics functioned in those regimes, what made some endure while others faltered, and what lessons follow for SAI.

\section{How Technical Metrics Have Enabled Multi-State Cooperation: Lessons from Existing Regimes}
\label{sec:precedents}

We do not attempt a comprehensive survey of international governance regimes; instead, we draw on a small set of illustrative precedents to show that, across otherwise very different domains, shared technical metrics and verifiable parameters have enabled effective approaches to international governance. Our claim is not that these patterns map onto SAI in every detail, nor that those enablers are always sufficient to ensure durable success. Rather, we argue from past experience that multi-state cooperation becomes feasible - and additional options are opened - when parties can anchor commitments in the type of technical building blocks we describe.

It is possible that a climate emergency will force states to attempt multi-state SAI coordination under pressure. A more constructive path is for willing states to develop the necessary technical infrastructure before any acute need arises. SAI is, of course, unique. But history shows that multi-state cooperation on other high-stakes or contested technologies has often been facilitated by the development of shared technical metrics and monitoring capabilities. In many cases, a measurable, independently verifiable quantity served as a catalyst that transformed political disagreements - even among adversarial states - into manageable coordination problems.

\noindent\textbf{Nuclear arms control (from SALT to New START, 1972 and on).}
Cold War nuclear arms control among rival superpowers would not have been possible without the technical infrastructure that addressed the ``trust but verify'' challenge. The regime operated bilaterally and reciprocally, without a strong central coordinator. It worked not because trust was manufactured diplomatically but because technical verification made it unnecessary as a precondition; each side could independently measure what mattered, like warhead counts, delivery systems, and testing activity. Nuclear arms control \cite{NewSTART}, from SALT in 1972 through New START in 2010, demonstrated that even adversarial states can accept quantitative limits on their most sensitive military assets when verification is technically credible, such as through satellite imagery, on-site inspections, and detailed data exchanges. For SAI, the lesson is that once the relevant variable can be independently measured, agreements can rest on what can be checked rather than on what has to be believed.

\noindent\textbf{The Open Skies Treaty (signed 1992, in force 2002).}
Although it has been hollowed out by the U.S. and Russian withdrawals, the Open Skies Treaty offers similar lessons. That regime, which still includes many European states and Canada, granted participating states reciprocal rights to conduct short-notice observation flights over one another's territory using agreed sensors and procedures, to monitor conventional military forces and activities \cite{aca_openskies}. Its purpose was not to resolve the underlying political conflict among participating states, but to reduce uncertainty and information asymmetries by creating a shared and accessible evidentiary record. For nearly two decades, rivals routinely inspected one another's sensitive activities under common technical protocols, demonstrating also that reciprocal observation can scale from bilateral arms-control arrangements to a multi-party verification regime. Parties could measure and verify the same reality using shared baselines and metrics - metrics that, importantly, did not rely on national disclosures. The treaty has since eroded, with U.S. withdrawal in 2020 and Russian withdrawal in 2021, but that erosion stemmed largely from broader geopolitical and compliance disputes rather than from any defects in the observation architecture itself. In the SAI context, SRF monitoring and particle traceability could play a similar role for the stratosphere to the Open Skies Treaty's agreed standards for sensors and common data record. And, whereas the Open Skies regime relies on episodic inspection, SRF monitoring and particle traceability would operate continuously.

\noindent\textbf{The Montreal Protocol (1987).}
The Montreal Protocol \cite{MontrealProtocol} is often cited as one of the most successful environmental regimes, succeeding in large part because it anchored compliance in directly measurable quantities - production and consumption of specific ozone-depleting substances - rather than in contested downstream metrics such as ozone-layer thickness. The ability to measure and verify these quantities at the national level was the technical enabler that made the regime politically possible. Operationally, each state regulates its own industries and reports annual production and consumption to the Ozone Secretariat (part of UNEP, based in Nairobi), which supports the Meetings of the Parties that review compliance and adjust commitments over time. Verification operates as a hybrid model: self-reported national data, cross-checked against trade statistics that identify discrepancies between imports and exports across borders, and reinforced by independent atmospheric measurements of ODS concentrations maintained by scientific networks. A dedicated Multilateral Fund financed developing countries' transitions, making compliance economically feasible, while trade restrictions on controlled substances with non-parties created strong incentives to join. For SAI, the Montreal Protocol illustrates what becomes achievable once a measurable control variable exists: built-in financing, trade-based enforcement, and institutional flexibility to update commitments as science evolves.

\noindent\textbf{The IAEA safeguards regime for civilian nuclear material (1970).}
The IAEA safeguards regime \cite{IAEA} may be the closest structural analog to what SAI governance would require, and it illustrates both how shared technical metrics enable multi-state coordination and where the approach has historically fallen short. For declared civilian nuclear materials, safeguards have worked well: unique identifiers, sealed containment, and routine inspections give the Agency extensive and ongoing visibility into the declared inventory, which is generally capable of detecting diversion. The regime has been weaker against undeclared activity - the cases of concern have typically been ones in which inspectors were denied access to suspect facilities, not ones in which signatures themselves failed. This is a crucial defect that SAI coordination can design around. Stratospheric aerosol would not be confined to sovereign facilities that can be closed to inspection; once injected, it would be globally detectable through stratospheric sampling and remote sensing. With particle traceability, the declared-vs-undeclared distinction becomes a property of the material itself - declared material carries a recognizable signature; any material without a declared signature could presumptively be treated as undeclared and subjected to forensic examination. The IAEA pattern is therefore worth emulating in its strengths (shared metrics, continuous technical picture, standing inspection machinery) and deliberately superseded in its core limitation (dependence on state-granted physical access).

\noindent\textbf{The International Telecommunication Union (est.\ 1865).}
The ITU \cite{ITU} is the oldest precedent examined here, and it demonstrates the durability of regimes built on directly measurable parameters. The ITU coordinates the use of the radio spectrum and, later, the geostationary orbit, by recording frequency and position assignments in a central registry. Operational responsibility is distributed: each state licenses and operates its own transmitters and satellites, while the ITU itself maintains the registry and hosts the periodic World Radiocommunication Conferences at which spectrum and orbital assignments are negotiated. The measurability of these parameters is the technical enabler that has allowed the regime to function for more than a century and a half. Physics also provides a natural enforcement mechanism: a state that ignores assigned slots might disrupt its own communications. For SAI, the measurability is analogous, but the self-defeating property is not. A state that over-injects might gain locally while imposing costs on others. Any coordination framework would therefore need monitoring, attribution, and audit mechanisms to create the incentives that physics provides naturally in the spectrum case.

\noindent\textbf{The UN Committee on the Peaceful Uses of Outer Space (COPUOS) (1959).}
COPUOS \cite{UNOOSA_COPUOS} is the durable UN forum behind the core space treaties and the UN registration system for objects launched into orbit \cite{UNOOSA_Registry}, anchoring the broadly accepted norm of self-declaration. Without an independent measurement layer, however, registration is delayed, function descriptions are vague, military payloads are under-described, and even the more recent Guidelines for the Long-term Sustainability of Outer Space Activities \cite{UNOOSA_LTS} remain voluntary norms without an independent verification mechanism. For SAI, COPUOS illustrates both the value of declaration norms and their limit: a registry without verification degrades into self-reporting whose accuracy cannot be checked.

\noindent\textbf{Maritime and aviation transponders: ADS-B (1944) and AIS (2002).}
Two transportation-related regimes illustrate how a simple technical identity standard, once adopted, can become the enforcement infrastructure on which multiple agreements rely. Aviation's Automatic Dependent Surveillance-Broadcast (ADS-B) plays an analogous role under the Chicago Convention framework and ICAO technical standards \cite{ICAO_Annex10}, with satellite-based receivers now providing global, independent observation that does not depend on ground-station coverage in any single state \cite{Aireon}. The Automatic Identification System (AIS), mandated under SOLAS Chapter V since 2002 \cite{IMO_AIS} and operating on ITU-allocated spectrum, was originally developed for collision avoidance between vessels at sea. It has since become central to enforcement across a range of regimes: the Port State Measures Agreement against illegal, unreported, and unregulated fishing \cite{FAO_PSMA,Pew_Oceans}; MARPOL pollution attribution; and UNCLOS flag-state responsibility. Those examples show the leverage of a shared technical identification: one measurement standard, originally adopted for safety, becomes the enabling infrastructure for later environmental and regulatory regimes. They also illustrate the failure mode that matters for SAI: vessels in the ``dark fleet'' switch off their transponders to evade sanctions or fishing rules \cite{Pew_Oceans}, and the intentional disabling of transponders remains a concern in both maritime and aviation monitoring. This case study highlights a crucial design advance of SAI particle traceability: the signature would be carried by the material itself, so unlike a transponder, it could not be switched off at the source.

\noindent\textbf{The Comprehensive Nuclear-Test-Ban Treaty (1996).}
The CTBT \cite{CTBT} demonstrates that a purpose-built global sensor network can serve as a powerful technical enabler of coordination, even if it also offers a cautionary lesson that effective measurement capabilities alone may not lead to a binding agreement. More than 300 seismic, hydroacoustic, infrasound, and radionuclide stations make nuclear-test violations detectable with high confidence: the International Monitoring System (IMS) identified all six North Korean tests, seismically and in several cases by radionuclide signatures as well. The coordination architecture is distinctive. IMS stations are hosted by national governments but operated to uniform international standards and feed raw data automatically to the International Data Center in Vienna, which the CTBTO Preparatory Commission runs and from which every signatory receives the full data stream. Most precedents discussed here verify self-reported national data - production volumes, monetary parities, treaty-limited systems - through inspection or audit. The CTBT does not: the IMS independently detects any test of significant yield, and that direct measurement is itself the primary evidence. The monitoring infrastructure is therefore a shared piece of global technical infrastructure rather than an audit layer sitting on top of national accounting - the closest architectural analog among these precedents to what Section~\ref{sec:building-blocks} proposes for SAI, where an atmospheric SRF monitoring network would directly observe the SAI-induced radiative effect rather than verify state-reported injection volumes. Yet despite near-universal support, the treaty has never entered into force because key states have not ratified it. The technical capability to enable cooperation exists; the political commitment to be bound by it does not yet fully exist. For SAI, this example underscores that technical enablers are necessary but not sufficient to overcome political barriers to coordination and mutual constraints.

\noindent\textbf{Lessons for a verifiable SAI coordination framework.}
These precedents converge on a consistent pattern: shared technical metrics and verification capabilities have been essential enablers of multi-state cooperation on high-stakes or contested technologies. Strong regimes - the ITU's frequency registry, the Montreal Protocol's production reporting backed by trade measures - have often paired an agreed technical variable with institutionally anchored verification. Other regimes have sometimes failed for diagnostic reasons: they lacked independent, real-time monitoring (IAEA before the Additional Protocol), they solved the measurement problem but could not close the political barriers to binding restrictions (CTBT), or political will eroded faster than the technical architecture could compensate (arms control).

On the whole, a key lesson is that measurability is often a precondition for cooperation. The capacity for independent observation and verification - rather than reliance on self-reporting or trust - makes agreements more possible and more durable. Those advantages are likely to be strongest when measurability operates through a common, open technical infrastructure that is easy to access and difficult to evade.

The practical implication for SAI is that the technical building blocks described in this paper - a shared control variable and independent monitoring with attribution - should be developed well before any possible SAI deployment. That is not because they guarantee cooperation, but because their absence would make cooperation, verification, and trust substantially harder. If and when a climate crisis forces states to the table, the question will be whether these enabling tools already exist or must be improvised under pressure.

\section{A Phased Approach: Test Before You Scale}
\label{sec:phased}

The technical capabilities described in this paper - and the multi-state coordination practices that would use them - cannot be adopted at full scale on day one. What follows is best understood as a research-governance rehearsal protocol rather than an operational governance proposal: a phased pathway for jointly developing and testing both, well before any decision to actually deploy SAI. As in the early phases of arms control, they would need a testing period in which measurement systems, attribution methods, and coordination procedures are validated, long before the stakes are high. This approach - starting with a small coalition and broadening as confidence in both the tools and the practices grows - has strong precedent in international environmental law \cite{Bodansky2024}, and criteria-gated progression from laboratory research through field testing to gradual deployment has been proposed specifically for SRM research \cite{Keith2014}.

Rather than waiting for a comprehensive international framework - which risks being either unachievable or counterproductive \cite{Biniaz2025,Reynolds2021} - willing states and research groups can begin building the technical capabilities and the multi-state coordination practices that would use them, in the polycentric pattern by which complex environmental governance has historically developed \cite{Ostrom2010}. Crucially, this means waiting on deployment, not waiting on the development of capabilities and practices: the technical infrastructure and institutional habits that any future arrangement might rely on can be built and validated at progressively larger scales while the question of whether to deploy remains open. A recent paper provides the technical basis for a gradual ramp-up \cite{Waxman2026}; here we describe coordination practices that could be tested alongside the technology.

Modern approaches to verification and coordination were rarely born as fully articulated governance systems. Often, the sequence ran in the opposite direction: a workable measurement capability emerged first, was exercised under real conditions, and only then became the basis for coordination or governance. National technical means in arms control followed this pattern most clearly: reconnaissance and sensing capabilities were developed as national intelligence tools and later protected in treaty language through the Anti-Ballistic Missile Treaty's recognition of national technical means of verification \cite{abm_treaty_1972}. The ITU monitoring system followed a more cooperative path, but the logic was similar: radio monitoring began as a practical service for managing interference, then matured into standardized reporting, designated monitoring stations, and formal regulatory rules under the Radio Regulations \cite{itu_monitoring_2014}. Maritime and aviation transponders followed the same pattern: AIS and ADS-B were adopted first as practical technical standards for collision avoidance, and the enforcement systems that now rely on them - SOLAS, the Port State Measures Agreement, MARPOL, UNCLOS, and the Chicago Convention framework - were layered onto a measurement standard already in routine operation \cite{IMO_AIS,FAO_PSMA,Pew_Oceans,ICAO_Annex10,Aireon}. Across these cases, the durable element was not regulation alone, but regulation anchored in a capability to measure \cite{ctbto_verification_regime,CTBT,iaea_safeguards,iaea_additional_protocol,montreal_article7,montreal_article6}.

That history supports the phased approach proposed here. For SAI, the relevant lesson is that we should not wait for a fully mature global framework before testing the technical foundations. Early phases should therefore focus on proving that the key observables - particle identity, attribution, radiative forcing, and deviation detection - can in fact be measured reliably and independently under realistic conditions. Coordination practices that leverage those capabilities can be developed in parallel to the demonstration.

As explained below, Phase~1 and Phase~2 should serve as technical prototyping; Phase~3 should serve as institutional rehearsal; and Phase~4 would be most effective only once both the measurement system and the governance routines have been validated. A phased pathway is therefore not merely a cautious implementation strategy. It is the historical pattern by which serious multi-state monitoring regimes have usually become credible \cite{itu_monitoring_2014,ctbto_verification_regime,CTBT,iaea_additional_protocol,montreal_article6}.

\noindent\textbf{Phase 1: Minuscule-scale trials.}
Experiments would be limited to minimal material quantities, with negligible environmental impact, and would comply with national environmental regulations (e.g., EU REACH, US EPA frameworks). Purpose: validate particle properties, traceability methods, and measurement systems \cite{RoyalSociety2025,COC2023,Waxman2026,Parson2013}.

\noindent\textbf{Phase 2: Bilateral or small-coalition trials.}
Two or three states would conduct limited, coordinated injection. The primary objective would not be to alter radiative forcing but to validate the technical infrastructure under real conditions. Participating states would deploy tagged particles at agreed injection points, operate independent monitoring stations, and attempt to recover and identify each other's particles from stratospheric samples. Success in this phase means demonstrating that tagged particles survive stratospheric transport and that attribution based on recovered particle signatures correctly identifies the injecting entity. This phase could also establish working relationships between national monitoring agencies and create institutional habits of data-sharing, joint reporting, and low-stakes dispute resolution. This phase would also provide the first opportunity to test the detection of untagged aerosol populations - particles lacking any participant's signature - which is the technical basis for identifying non-member injection programs in later phases.

\noindent\textbf{Phase 3: Expanding participation and testing governance mechanisms.}
In the third phase, participation would expand to include a wider set of states and entities with possibly differing climate interests. Injection volumes would increase but remain below the threshold at which forcing effects may cause detectable climate impacts. The purpose of this phase would be to rehearse the governance machinery: the process by which forcing envelopes would be negotiated, the procedures through which detected deviations would be reported and adjudicated. Simulated adversarial exercises (see Section~\ref{sec:scenarios}), in which one participant deliberately deviates from its committed envelope by a pre-announced amount, would allow the framework to test its detection, attribution, and enforcement pipelines end-to-end. This phase would also be the natural point at which the coalition builds operational architectures towards Phase~4.

\noindent\textbf{Phase 4: Large-scale testing or pilot ramp-up under full governance.}
Experiments at this scale may have detectable effects on the climate. If they took place, they should be government-led or led by appropriate international bodies. No large-scale testing should proceed without (a) demonstrated safety across all prior phases, and (b) governance arrangements providing coordination authority for this scale. By the time Phase~4 begins, the operator(s) should be fully constituted: chartered, staffed, funded, procurement-capable, and accountable to the authorizing states. At this point, the monitoring infrastructure would have been validated and the governance procedures exercised.

\begin{figure*}[ht]
\centering
\resizebox{0.85\textwidth}{!}{
\begin{tikzpicture}[
    >=Stealth,
    node distance=0.5cm,
    phasebox/.style={
        rectangle,
        draw=prlblue,
        line width=0.55pt,
        fill=prlbluetint,
        minimum width=3.8cm,
        minimum height=2.8cm,
        align=center,
        inner sep=8pt,
        font=\small,
        rounded corners=3pt,
    },
    arr/.style={
        -{Stealth[length=6pt, width=4pt]},
        line width=1.0pt,
        prlblue,
    },
]

\node[phasebox] (p1) {%
    \textbf{\color{prlblue}Phase 1}\\[4pt]
    Minuscule-scale trials\\[2pt]
    {\scriptsize\color{prllg}Validate particles,}\\
    {\scriptsize\color{prllg}traceability, measurements}};

\node[phasebox, right=1.2cm of p1] (p2) {%
    \textbf{\color{prlblue}Phase 2}\\[4pt]
    Bilateral trials\\[2pt]
    {\scriptsize\color{prllg}Tagged particle recovery,}\\
    {\scriptsize\color{prllg}attribution validation}};

\node[phasebox, right=1.2cm of p2] (p3) {%
    \textbf{\color{prlblue}Phase 3}\\[4pt]
    Expanded coalition\\[2pt]
    {\scriptsize\color{prllg}Governance machinery,}\\
    {\scriptsize\color{prllg}adversarial exercises}};

\node[phasebox, right=1.2cm of p3] (p4) {%
    \textbf{\color{prlblue}Phase 4}\\[4pt]
    Full governance\\[2pt]
    {\scriptsize\color{prllg}Large-scale pilot,}\\
    {\scriptsize\color{prllg}detectable climate effects}};

\draw[arr] (p1) -- (p2);
\draw[arr] (p2) -- (p3);
\draw[arr] (p3) -- (p4);

\node[below=0.6cm of p1, font=\scriptsize\color{prlgray}, align=center] {%
    Technical\\infrastructure};
\node[below=0.6cm of p4, font=\scriptsize\color{prlgray}, align=center] {%
    Coordination\\practices};

\draw[-{Stealth[length=5pt, width=3pt]}, line width=0.8pt, prlgray]
    ([yshift=-1.6cm]p1.south) -- ([yshift=-1.6cm]p4.south)
    node[midway, below=3pt, font=\scriptsize\bfseries\color{prlgray}]
    {Criteria-gated progression};

\end{tikzpicture}}
\caption{Phased co-development of technical infrastructure and the multi-state coordination practices that would use it, from minuscule-scale trials through full operational scale. Technical infrastructure (monitoring, detection, traceability, attribution) is prototyped and validated in the early phases, providing the foundation on which coordination practices (forcing envelopes, audit procedures, reserves, escalation, response authority) are progressively built in later phases. Progression is criteria-gated rather than calendar-gated: each transition requires demonstrated technical and institutional readiness from the prior phase.}
\label{fig:phased}
\end{figure*}
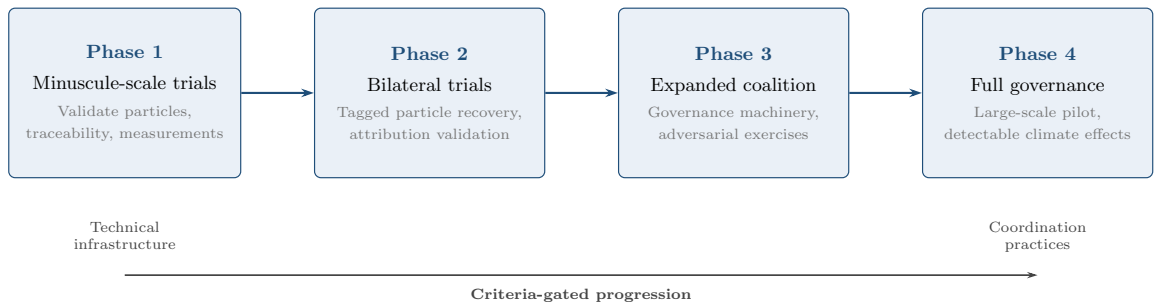

Progression between phases should be criteria-gated, not calendar-gated. This applies to governance readiness as much as to technical readiness: a state or coalition could not enter Phase~4 without the monitoring infrastructure being validated and the institutional arrangements being in place. Failure at any phase should halt progression until the deficiency is resolved.

\section{Illustrative Scenarios}
\label{sec:scenarios}

The scenarios below illustrate how some of the multi-state SAI coordination challenges discussed in Section~\ref{sec:challenges} could play out in practice, and how the technical building blocks could provide the data needed to address them. The scenarios deliberately consider strategic, deceptive, or reckless actors; they are illustrative, not exhaustive.

\subsection{How would an SAI program work in practice?}

Assume the phased trajectory of Section~\ref{sec:phased} has played out: technical tools validated, governance machinery rehearsed, and coordination scaled to a coalition with the industrial capacity and stratospheric-delivery infrastructure required to run a program at planetary scale. Only a small number of states are plausible founders - those with both the aerospace base to sustain continuous injection and the diplomatic weight to bring others under a common envelope \cite{Horton2025_WhoDeploy,Nielsen2025,Abatayo2020,Ramge2025}. Other multilateral configurations are possible; the scenarios below use a US-China core, open to accession, set up as just one example.

Within that world, imagine that a coalition executive body sets and refreshes a forcing envelope, measured by a distributed atmospheric SRF monitoring network whose data is publicly accessible and open to independent interrogation. Particle production is spread across factories in several participating states, with tagged-material reserves prepositioned to mitigate disruption. Injection flights operate under either a multi-operator arrangement - participating air forces flying the mission under joint authorization - or a single-operator arrangement in which states jointly authorize a mandated entity. In either configuration, the two technical building blocks (SRF as control variable, particle traceability for attribution) underpin compliance, attribution, audit, and continuity through the same observable data.

This is a world in which the two technical building blocks introduced in Section~\ref{sec:building-blocks} and the governance capabilities rehearsed through the phases in Section~\ref{sec:phased} have matured into routine operation. The framework is imperfect - accession is uneven, enforcement procedures are still maturing, and not every state is fully aligned with the program's goals. But SRF is measured and published, particle traceability works, and when something goes wrong - as it does in the scenarios that follow - the coalition has the tools, habits, and authority to see it and respond.

\subsection{Adversarial Scenarios}

These scenarios are primarily framed for the multi-operator case, in which several states each conduct injections under shared rules. Under a single-operator architecture, analogous adversarial dynamics take the form of operator deviation from the joint mandate (rather than between-state cheating), member-state under-funding or withholding of reserve capacity (rather than under-injection), and coalition-level coercion through threatened withdrawal of political or financial support (rather than threatened cessation of injection). The technical building blocks - SRF monitoring and particle traceability, as well as shared data - provide the same evidentiary foundation in both cases; what changes is that auditability runs through the operator's books and the principal-agent relationship between operator and coalition rather than between member states.

\noindent\textbf{Covert deviation from the committed envelope.}

\emph{Scenario.} A participating state secretly deviates from its committed injection envelope - either exceeding it, to cool its own regions at the expense of altered precipitation patterns elsewhere, or falling below it, relying on other participants to maintain the aggregate forcing target while saving costs.

\emph{Response.} Continuous SRF monitoring detects the regional anomaly: observed forcing in the relevant latitude band departs from the sum of all committed envelopes. Stratospheric aerosol sampling then recovers tagged particles in counts inconsistent with the suspected state's declared program - an excess in the over-injection case, a shortfall in the under-injection case. The combination of forcing anomaly and tagged-particle mismatch identifies both the fact and the perpetrator. Because the evidence is physical, the offending state cannot plausibly attribute the deviation to measurement error, and the coordination framework can respond on a factual basis that is difficult to dispute. Under a single-operator architecture, member-state underfunding may not appear in atmospheric data alone (the operator injects as a single stream); detection runs through corroborating financial and reserve audits.

\noindent\textbf{Non-member injection programs.}

\emph{Scenario.} A state or a group of states outside the coalition initiates its own injection program, ranging from covert small-scale activity to a full-scale rival program with different climate objectives - for example, heavy injection at latitudes that benefit its own climate at the expense of equatorial states inside the framework.

\emph{Response.} Routine stratospheric sampling detects aerosol populations without any participating operator's declared chemical or isotopic signature; these may be untagged, or carry a different signature from the rival program. SRF monitoring quantifies the SAI-induced forcing contribution of these non-member aerosols by latitude and time. The dual signal - an unexplained SRF deviation plus the presence of foreign-signatured particles - constitutes the strongest evidence the architecture can provide. In the case of a fully developed rival program, the coordination framework may be able to characterize the non-member's activity in detail (injection locations, rates, seasonal patterns) without requiring any cooperation, and may adjust its own injection accordingly. In the case of small-scale covert activity, the signal narrows the set of possible sources but, by itself, does not identify the actor; attribution then depends on whatever forensic identification, intelligence, or diplomatic work the mandated authority is empowered to conduct. In either case, the architecture frames the problem - there is untagged or differently-tagged material altering the global radiation balance - and provides the evidentiary basis for whatever response participating states may take. How they choose to respond - through diplomatic pressure, trade measures, or other tools - lies outside this paper's scope. Critically, it is non-trivial for a non-member to remain covert at any meaningful scale: particles lacking the coalition's signature would betray the activity to any coalition state with sampling capability.

\noindent\textbf{Fabricated breach claims.}

\emph{Scenario.} Assuming an extensive international agreement is in place at a certain future time\footnote{Intermediate scenarios where an agreement covers only a group of coordinating states are outside the framework of this paper.}, a participating state alleges that another member's over-injection caused a severe drought, seeking remedy or political concessions.

\emph{Response.} Within the framework's audit mechanism, claims are anchored in measured forcing deviations, not in disputed downstream harm. The first question is factual: did the accused state violate its forcing envelope? SRF monitoring provides the answer. If no envelope violation occurred, the claim lacks a basis under the agreement's rules, regardless of whether a drought actually occurred. If a violation did occur, particle-tagging data confirms the responsible party and the extent of the violation. Any finding is anchored in the measured deviation in forcing, not in the claimant's assessment of downstream impacts. This structure does not prevent states from raising legitimate grievances; it makes it difficult to sustain fabricated or exaggerated claims because the evidentiary standard is observational rather than model-dependent.

\section{Conclusion}
\label{sec:conclusion}

This paper argues that the political viability of any future multi-state SAI coordination framework could be strengthened by two technical building blocks: SAI-induced radiative forcing as a shared, operator-independent control variable, and particle traceability as an intrinsic attribution mechanism, each feeding a shared, publicly accessible monitoring database that any signatory can independently interrogate. Without these capabilities, many coordination challenges - such as reconciling national interests, ongoing verification, distinguishing contributions, attribution and auditing, and operational continuity - are hard to address solely through political means. With these suggested technical capabilities, each of the described challenges has a measurable technical basis. That technical basis would not eliminate political disagreement but would anchor it in observable facts rather than contested claims.

Figure~\ref{fig:phased} shows the developmental axis - how the technical building blocks could mature alongside the governance-related toolbox. Table~\ref{tab:summary} summarizes some of the multi-state SAI governance challenges introduced in Section~\ref{sec:challenges} and the technical building blocks that could help address them.

\begin{table*}[ht]
\centering\small
\caption{Examples of multi-state SAI governance challenges and the technical building blocks that could support better resolution. The list is illustrative, not exhaustive; the building blocks are technical inputs, not a complete governance system.}
\label{tab:summary}
\renewcommand{\arraystretch}{1.2}
\begin{tabular}{p{5.0cm}p{8.0cm}}
\toprule
\textbf{Governance challenge} & \textbf{Technical building blocks that could help} \\
\midrule
Reconciling national interests & SRF envelopes as shared control variable + particle traceability \\
Ongoing verification & Continuous SRF monitoring + particle traceability \\
Distinguishing contributions & Particle traceability (internal) + untagged-aerosol detection (external) \\
Attribution and auditing & SRF envelope as observational benchmark \\
Operational continuity & SRF early warning + tagged-particle gap attribution \\
\bottomrule
\end{tabular}
\end{table*}

The alternative to beginning this work now is either a long delay or rushed improvisation under pressure. If a climate emergency ever forces states to the table, the tools for auditable coordination will either already exist or will have to be assembled against a running clock. The historical record shows that credible multi-state cooperation on contested technologies has rested on measurement and verification capabilities in place to support it - and has faltered where those capabilities were absent or incomplete. SAI is unlikely to be an exception. The broader climate governance literature has reached a similar conclusion through a different route: that effective response to complex, multi-scale problems comes not from a single comprehensive treaty but from polycentric arrangements that build outward from smaller, willing coalitions \cite{Ostrom2010, Victor2011, KeohaneVictor2016}.

We are explicit about what this paper does not do. It does not propose the institutional, legal, or political arrangements that a functioning governance system would require. Those arrangements - and the expertise needed to design them - belong to communities better placed than we are. What the paper offers is the evidentiary foundation on which any such arrangements could rest. Whatever institutional form an eventual governance system might take, it would be more workable if the measurements underneath it exist, are independently verified, and are trusted.

The phased pathway described in Section~\ref{sec:phased} is open to any coalition of states or research groups willing to begin. Early phases would require no treaty - only the discipline to measure what is measurable and to publish what is observed in minuscule scales that pose no environmental harm. That discipline, accumulated over progressively expanding trials, is what would allow a future framework - if one ever becomes necessary - to be built from capabilities already validated, rather than from promises yet to be kept. In that sense, the path from particles to policy will run not only on politics, but on measurement and observation as well.

\section*{Declarations}
Matthew C.~Waxman consults for Stardust on international law and technology ethics matters. Stardust is a company researching stratospheric aerosol injection technologies.

\bibliography{refs-bibfile}

\appendix

\section{Physical and Technical Basis}
\label{app:physical-basis}

This Annex provides order-of-magnitude estimates underlying the claims in the main text. The purpose is to establish scale rather than precise design values. While all quantities depend on particle properties, injection strategy, and the chosen SRF envelope, the relevant magnitudes are robust: climate-relevant SAI requires Mt/yr material flows\footnote{``Mt'' denotes one million metric tons, equivalent to \(1~\mathrm{Tg}\).}, stratospheric residence times of order one year, and atmospheric transport that erases simple source attribution unless particles carry intrinsic identity.

\begin{table}[!htbp]
\centering\small
\caption{Representative physical and operational scales.}
\renewcommand{\arraystretch}{1.1}
\begin{tabular}{p{6cm}p{5cm}}
\toprule
\textbf{Quantity} & \textbf{Scale} \\
\midrule
Excess GHG RF compared to 1850 & \(\sim3~\mathrm{W\,m^{-2}}\) \\
SRF per mass & \(\sim0.15~\mathrm{W\,m^{-2}/Mt}\) \\
Stratospheric burden & \(\sim17~\mathrm{Mt}\) for \(\sim2.5~\mathrm{W\,m^{-2}}\) \\
Annual release & \(17~\mathrm{Mt\,yr^{-1}}\) \\
Particle size & peak \(\sim0.5~\mu\mathrm{m}\) \\
Injection altitude & \(\sim18.5~\mathrm{km}\) \\
Residence time & \(1~\mathrm{yr}\) \\
Zonal mixing & weeks-months \\
Meridional mixing & \(\sim10^3~\mathrm{km/month}\) \\
Fleet scale & \(2000\) aircrafts \\
Annual recurring cost &  \$100 B\\
\bottomrule
\end{tabular}
\end{table}

The global-mean incoming solar flux is \(\simeq 340~\mathrm{W\,m^{-2}}\), so reflecting \(1\%\) corresponds to \(\sim 3.4~\mathrm{W\,m^{-2}}\), comparable to excess GHG RF compared to 1850 (\(\sim 3~\mathrm{W\,m^{-2}}\)). A practical scaling between SAI-induced radiative forcing and stratospheric particle burden is \cite{Stardust2026Yoav}
\[
M_{\rm burden} \simeq 7~\mathrm{Mt}
\left(\frac{{\rm SRF}}{1~\mathrm{W\,m^{-2}}}\right),
\]
implying that \(\sim 2.5~\mathrm{W\,m^{-2}}\) requires a burden of order \(17~\mathrm{Mt}\). For a residence time \(\tau_s \sim 1~\mathrm{yr}\), the corresponding replenishment rate is
\[
\dot{M} \simeq \frac{M_{\rm burden}}{\tau_s} \sim 17~\mathrm{Mt\,yr^{-1}},
\]
while an intermediate objective (\(\sim 0.7~\mathrm{W\,m^{-2}}\)) corresponds to \(\sim 5~\mathrm{Mt\,yr^{-1}}\).

Transport in the lower stratosphere is characterized by rapid zonal mixing and slower meridional spreading. Zonal winds \(u_z\sim10~\mathrm{m\,s^{-1}}\) imply longitudinal homogenization on timescales of weeks to a few months, while meridional transport can be approximated, after averaging over the three-dimensional flow, by an effective (eddy) diffusivity \(K\sim10^6~\mathrm{m^2\,s^{-1}}\) on monthly scales \cite{Haynes2000_EffectiveDiffusivity1,Haynes2000_EffectiveDiffusivity2,Butchart2014,SPARC2006_ASAP}, giving a characteristic scale
\[
L_{\rm diff} \simeq (Kt)^{1/2} \sim 10^3~\mathrm{km}
\]
over one month. These scales define the natural resolution for controllability and verification.

The deployment system implied by these requirements is large but bounded. Assuming \(\sim10~\mathrm{kton\,yr^{-1}}\) per aircraft \cite{Stardust2026Composite},
\[
N_{\rm aircraft} \sim 500\left(\frac{\dot{M}}{5~\mathrm{Mt\,yr^{-1}}}\right),
\]
rising to \(\sim2000\) at full scale. The total cost can be expressed as
\[
C \sim c_{\rm mat}\dot{M}\sim \left(\frac{c_{\rm mat}}{\$5\,\mathrm{kg}^{-1}}\right)\left(\frac{\dot{M}}{5~\mathrm{Mt\,yr^{-1}}}\right)\sim\$25~\mathrm{B\,yr^{-1}},
\]
where \(c_{\rm mat}\) represents the particle costs (assuming recurring cost is dominated by particle manufacturing), rising to \(\sim\$100~\mathrm{B\,yr^{-1}}\) at full scale.

Particle tagging reduces attribution to an encoding problem. Assigning each batch a unique identifier, the collision probability is
\[
P_{\rm coll} \simeq \frac{N(N-1)}{2^{b+1}},
\]
 requiring
\[
b \gtrsim 2\log_2 N.
\]
With \(N\sim10^7\text{--}10^8\) (for a deployment rate of \(5\text{--}10~\mathrm{Mt\,yr^{-1}}\), combined with batch sizes of order \(1\text{--}10~\mathrm{t}\)), one obtains \(b\sim50\text{--}60\) bits; a practical design therefore uses \(64\text{--}96\) bits. For example, an implementation with \(m\) independent trace-element or isotopic marker channels, each controlled at \(q\) distinguishable abundance levels, encodes \(b=m\log_2 q\) bits; thus \(m\simeq20\) channels with \(q=10\) levels provide \(\simeq66\) bits, while \(m\simeq30\) provide \(\simeq100\) bits, before adding redundancy for measurement noise.

Monitoring separates the measurement of the aerosol layer from the measurement of the resulting radiation field. Aerosol-based methods directly observe the controlled quantity, allowing SRF to be inferred from particle burden and optical properties. Solar occultation systems achieve \(\sim5\%\) accuracy with \(\sim0.5~\mathrm{km}\) vertical resolution and can detect perturbations of order \(10^{-4}\text{--}10^{-3}\) against a background optical depth of \(\sim10^{-3}\) \cite{SPARC2006_ASAP,sageiii_atbd}, at system costs of order \(\$150\text{--}300~\mathrm{M}\). TOA flux measurements provide a complementary radiation budget closure measurement, but are not the primary compliance observable because they require extracting a small perturbation (\(\sim0.1\text{--}1~\mathrm{W\,m^{-2}}\)) from a large and variable background (\(\sim100\text{--}250~\mathrm{W\,m^{-2}}\)), with regional uncertainties of \(\sim2\text{--}3~\mathrm{W\,m^{-2}}\), detection thresholds \(\gtrsim0.7~\mathrm{W\,m^{-2}}\), and system costs of order \(\$1\text{--}3~\mathrm{B}\).

These same estimates also explain why the architecture assumed here is better matched to manufactured solid aerosols than to \(\mathrm{SO_2}\) precursor injection. With \(\mathrm{SO_2}\), the injected material is not the final radiatively active particle: sulfate aerosols form later through chemistry and microphysics, so their size distribution, optical properties, surface area, residence time, and SRF depend on local atmospheric conditions and background aerosol. This weakens controllability and raises safety concerns, including enhanced heterogeneous chemistry relevant to ozone depletion and stratospheric heating from infrared absorption. Manufactured solid particles can instead be specified before release -- size, composition, optical properties, surface chemistry, and traceable identity -- and the requirement that they remain chemically inert and stable in the stratosphere is what preserves both safety and controllability.

The physical scales in this Annex therefore point to the governance architecture used in the main text. SRF is the highest-level quantity that is both SAI-specific and independently quantifiable from direct aerosol-layer measurements, making it the natural control variable. Atmospheric transport sets the relevant verification scale, favoring monthly and \(\sim10^3~\mathrm{km}\) averages rather than finer operational grids. Particle tagging is needed because mixing makes aggregate aerosol measurements insufficient for attribution, while tags convert the aerosol field into an auditable inventory. Finally, the mass, fleet, and cost scales place SAI firmly in the domain of state-level infrastructure, where verification, coordination, and long-term operational continuity are feasible but unavoidable requirements.

\end{document}